\documentclass[twocolumn,showpacs,aps,prl,superscriptaddress]{revtex4}

\usepackage{graphicx}
\usepackage{dcolumn}
\usepackage{amsmath}
\usepackage{epsfig}
\usepackage{color}
\usepackage{url}

\input babarsym

\newcommand{\BABARPubYear}    {10}
\newcommand{\BABARPubNumber}  {032}
\newcommand{\SLACPubNumber} {14378}
\newcommand{\LANLNumber} {1102.4565}

\begin{document}

\preprint{\babar-PUB-\BABARPubYear/\BABARPubNumber}
\preprint{SLAC-PUB-\SLACPubNumber}
\begin{flushleft}
\babar-PUB-\BABARPubYear/\BABARPubNumber\\
SLAC-PUB-\SLACPubNumber\\
arXiv:\LANLNumber\ [hep-ex]\\[10mm]
\end{flushleft}

\title{
{\large \bf
Evidence for the $\boldmath h_b(1P)$ meson in the decay $\boldmath \Upsilon(3S)\rightarrow \pi^0 h_b(1P)$ 
 }
}

%
\author{J.~P.~Lees}
\author{V.~Poireau}
\author{E.~Prencipe}
\author{V.~Tisserand}
\affiliation{Laboratoire d'Annecy-le-Vieux de Physique des Particules (LAPP), Universit\'e de Savoie, CNRS/IN2P3,  F-74941 Annecy-Le-Vieux, France}
\author{J.~Garra~Tico}
\author{E.~Grauges}
\affiliation{Universitat de Barcelona, Facultat de Fisica, Departament ECM, E-08028 Barcelona, Spain }
\author{M.~Martinelli$^{ab}$}
\author{D.~A.~Milanes$^{ab}$ }
\author{A.~Palano$^{ab}$ }
\author{M.~Pappagallo$^{ab}$ }
\affiliation{INFN Sezione di Bari$^{a}$; Dipartimento di Fisica, Universit\`a di Bari$^{b}$, I-70126 Bari, Italy }
% ...
\collaboration{The \babar\ Collaboration}
\noaffiliation

\date{\today}

\begin{abstract}
Using a sample of 122 million $\Upsilon(3S)$ events recorded with the \babar\ detector at the PEP-II
asymmetric-energy $e^+e^-$ collider at SLAC, we search for the $h_b(1P)$ spin-singlet partner of the
$P$-wave $\chi_{bJ}(1P)$ states in the sequential decay $\Upsilon(3S)\rightarrow \pi^0 h_b(1P),$ $h_b(1P)\rightarrow \gamma \eta_b(1S)$.
We observe an excess of events above background in the distribution of the recoil mass against the 
$\pi^0$ at mass $9902\pm 4 {\rm (stat.)}\pm 2 {\rm (syst.)}$ MeV/$c^2$.  
The width of the observed signal is 
consistent with experimental resolution, and its significance is 3.1$\sigma$, including systematic uncertainties.
We obtain the value $(4.3 \pm 1.1$ (stat.) $\pm 0.9$ (syst.))$\times 10^{-4}$ for the product branching fraction
${\mathcal B}(\Upsilon(3S)\rightarrow \pi^0 h_b)\times {\mathcal B}(h_b\rightarrow \gamma \eta_b)$. 

\end{abstract}


\maketitle

To understand the spin dependence of $q \bar{q}$ potentials for heavy quarks,
it is essential to measure the hyperfine mass splitting for $P$-wave states.
In the non-relativistic approximation, the hyperfine splitting is proportional to the square of the wave function at the origin,
which is expected to be non-zero only for $L=0$, 
where $L$ is the orbital angular momentum of the $q\bar{q}$ system.  
For $L=1$, the splitting between the spin-singlet (${}^1P_1$) and the spin-averaged triplet state ($\langle {}^3P_J\rangle$)
is expected to be $\Delta M_{\rm HF}=M(\langle{}^3P_J\rangle)-M({}^1P_1)\sim 0$.
The ${}^1 P_1$ state of  bottomonium, the $h_b(1P)$, 
is the axial vector partner of the $P$-wave $\chi_{bJ}(1P)$ states.
Its expected mass, computed as the spin-weighted center of gravity of the $\chi_{bJ}(1P)$ states, is 9899.87$\pm$ 0.27 MeV/$c^2$~\cite{ref:PDG}.
Higher-order corrections might cause a small deviation from this value, but a
hyperfine splitting larger than 1 MeV/$c^2$ might be
indicative of a vector component in the confinement potential~\cite{ref:Rosner2002}.
The hyperfine splitting for the charmonium ${}^1 P_1$ state $h_c$ is measured by the BES and CLEO experiments~\cite{ref:BEShc,ref:CLEOhc0,ref:CLEOhc}
to be $\sim$0.1 MeV/$c^2$.
An even smaller splitting is expected for the much heavier bottomonium system~\cite{ref:Rosner2002}.

The $h_b(1P)$ state is expected to be produced in $\Upsilon(3S)$ decay via 
$\pi^0$ or di-pion emission, and to undergo a subsequent $E1$
transition to the $\eta_b(1S)$, with branching fraction (BF) 
${\mathcal B}(h_b(1P)\rightarrow \gamma \eta_b(1S))\sim (40-50)\%$~\cite{ref:Rosner2002,ref:Rosner2005}.
The isospin-violating  
decay $\Upsilon(3S) \rightarrow \pi^0 h_b(1P)$ is expected to have a BF of about 0.1\%~\cite{ref:BFpredict,ref:Godfrey}, while 
theoretical predictions for the transition $\Upsilon(3S) \rightarrow \pi^+\pi^- h_b(1P)$ 
range from $\sim 10^{-4}$~\cite{ref:BFpredict} up to $\sim 10^{-3}$~\cite{ref:ratiopredict}.   
A search for the latter decay process in \babar\ data yielded an upper limit on the BF of $1.2\times 10^{-4}$ at 90\% confidence level (C.L.)~\cite{ref:simone}.  
The CLEO experiment reported the 90\% C.L.
limit ${\mathcal B}(\Upsilon(3S)\rightarrow \pi^0 h_b(1P))<0.27\%$, assuming the mass of the $h_b$ to be 9900 MeV/$c^2$~\cite{ref:BFcleo}. 

In this paper, we report evidence for the $h_b(1P)$ state in the decay  
$\Upsilon(3S)\rightarrow \pi^{0} h_b(1P)$.   
The data sample used was collected with the \babar\ detector~\cite{ref:babar}
at the PEP-II asymmetric-energy $e^+e^-$ collider at SLAC, and corresponds to 
28 \invfb of integrated luminosity at a center-of-mass (CM) 
energy of 10.355~GeV, the mass of the $\Upsilon(3S)$ resonance. This sample contains $(122 \pm 1)$ million $\Upsilon(3S)$ events.
Detailed Monte Carlo (MC) simulations~\cite{ref:ftn1} of samples of  
exclusive $\Upsilon(3S)\rightarrow \pi^0 h_b(1P),$ $h_b(1P)\rightarrow \gamma \eta_b(1S)$ decays 
(where the $h_b(1P)$ and $\eta_b(1S)$ are hereafter referred to as the $h_b$ and the $\eta_b$), 
and of inclusive $\Upsilon(3S)$ decays, 
are used in this study.  These samples correspond to 34,000 signal and 215 million $\Upsilon(3S)$ events, respectively.  
In the inclusive $\Upsilon(3S)$ MC sample a BF of 0.1\% is assumed for the decay $\Upsilon(3S)\rightarrow \pi^0 h_b$~\cite{ref:BFpredict}.

The trajectories of charged particles are reconstructed
using a combination of five layers of double-sided silicon strip
detectors and a 40-layer drift chamber, both operating inside the 1.5-T magnetic field
of a superconducting solenoid.
Photons are detected, and their energies measured, with a CsI(Tl) electromagnetic calorimeter
(EMC), also located inside the solenoid.
The \babar\ detector is described in detail elsewhere~\cite{ref:babar}.

The signal for $\Upsilon(3S)\rightarrow \pi^{0} h_b$ decays is extracted
 from a fit to the inclusive recoil mass distribution against the $\pi^0$ 
candidates ($m_\mathrm{recoil}(\pi^0$)).   
It is expected to appear as a small excess centered near 9.9 GeV/$c^{2}$ on top of the very large non-peaking background produced from
 continuum events ($e^+e^- \rightarrow q\bar q$ with $q=u,d,s,c$) and
 bottomonium decays.  
The recoil mass, $m_\mathrm{recoil}(\pi^0)=\sqrt{(E^*_\mathrm{beam}-E^{*}(\pi^0))^2-p^*(\pi^0)^2}$, 
where $E^*_\mathrm{beam}$ is the total beam CM energy, and $E^{*}(\pi^0)$ and $p^*(\pi^0)$ are 
the energy and momentum of the $\pi^0$, respectively, computed in the $e^+e^-$ CM frame (denoted by the asterisk).
The search for an $h_b$ signal, requiring detection only of the recoil $\pi^0$, proved 
unfruitful because of the extremely large 
associated $\pi^0$ background encountered.  In order to reduce this background significantly, we exploit the fact that the $h_b$ should 
decay about half of the time~\cite{ref:Rosner2002, ref:Rosner2005} to $\gamma\eta_b$, and so require in addition the detection of a 
photon consistent with this decay.  
The precise measurement of the $\eta_b$ mass~\cite{ref:etabdiscovery} defines a restricted energy range for 
a photon candidate compatible with this subsequent $h_b$ decay.  
The resulting decrease in $h_b$ signal efficiency is offset by reduction of the $\pi^0$ background by a factor of about twenty.  
A similar approach led to the observation by CLEO-c, and then by BES, of the $h_c$ in the decay chain $\psi(2S)\to h_c\pi^0\to
\eta_c\gamma\pi^0$~\cite{ref:BEShc,ref:CLEOhc0,ref:CLEOhc}, where the $\eta_c$ was identified both exclusively
(by reconstructing a large number of hadronic modes) and inclusively.

The signal photon from $h_b\rightarrow \gamma \eta_b$ decay is monochromatic in the $h_b$ rest-frame and is expected to peak at $\sim$490 MeV in the $e^+e^-$ CM frame, 
with a small Doppler broadening that arises from the motion of the $h_b$ in that frame; 
the corresponding energy resolution is expected to be $\sim 25$ MeV.  The Doppler broadening is negligible compared with the energy resolution.       
Figure~1 shows the reconstructed CM energy distribution of candidate photons in the region 250-1000 MeV for simulated 
$\Upsilon(3S)\rightarrow \pi^0 h_b,$ $h_b\rightarrow \gamma \eta_b$ events before the application of selection criteria; 
the signal photon from $h_b\rightarrow \gamma \eta_b$ decay appears as a peak on top of a smooth background.  
We select signal photon candidates with CM energy in the range 420-540 MeV (indicated by the shaded region in Fig.~1).

\begin{figure}[ht!]
\begin{center}
 \includegraphics[width=.45\textwidth]{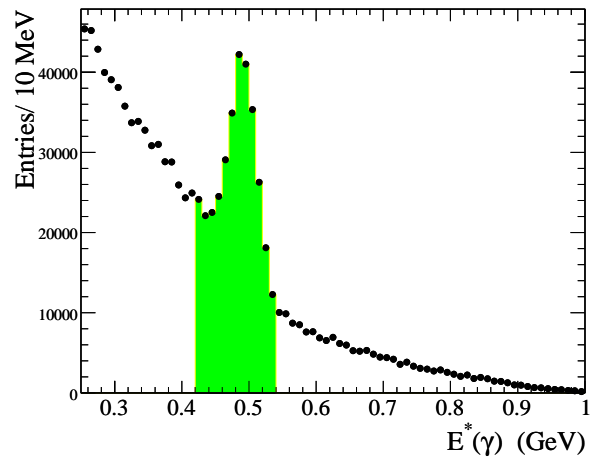}
 \begin{picture}(0.,0.)
    \end{picture}
  \caption{The reconstructed CM energy distribution of the candidate photon in
simulated $\Upsilon(3S)\rightarrow \pi^0 h_b(1P),$ $h_b(1P)\rightarrow \gamma \eta_b(1S)$ events.  The shaded region 
indicates the selected $E^*(\gamma)$ signal region. }
\end{center}
\end{figure}

We employ a simple set of selection criteria to suppress backgrounds
while retaining a high signal efficiency.
These selection criteria are chosen by optimizing
the ratio of the expected signal yield to the square root of the background. 
The $\Upsilon(3S)\rightarrow \pi^0 h_b$, $h_b\rightarrow \gamma \eta_b$ MC signal sample is used in the optimization, while a 
small fraction (9\%) of the total data sample is used to model the background.  
We estimate the background contribution in the signal region, defined by $9.85<m_\mathrm{recoil}(\pi^0)<9.95$ GeV/$c^2$, 
using the sidebands of the expected $h_b$ signal region, 
$9.80<m_\mathrm{recoil}(\pi^0)<9.85$ GeV/$c^2$ and $9.95<m_\mathrm{recoil}(\pi^0)<10.00$ GeV/$c^2$.  

The decay of the $\eta_{b}$ is expected to result in high final-state track multiplicity.  
Therefore, we select a hadronic event candidate by requiring that it have at least four charged-particle tracks and
a ratio of the second to zeroth Fox-Wolfram moments~\cite{ref:fox} less than 0.6~\cite{ref:r2}.  

For a given event, we require that the well-reconstructed tracks yield a successful fit to a primary vertex 
within the $e^+e^-$ collision region.  We then 
constrain the candidate photons in that event to originate from that vertex. 

A photon candidate is required to deposit a minimum energy in the laboratory frame of 50 MeV into a contiguous EMC crystal cluster that is  
isolated from all charged-particle tracks in that event.  
To ensure that the cluster shape is consistent with that for an electromagnetic shower,
its lateral moment~\cite{ref:LAT} is required to be less than 0.6.

A $\pi^0$ candidate is reconstructed as a photon pair with invariant mass $m(\gamma \gamma)$ in the range 55--200 MeV/$c^2$ (see Fig.~2).  
In the calculation of $m_\mathrm{recoil}(\pi^0)$, the $\gamma$-pair invariant mass is 
constrained to the nominal $\pi^0$ value~\cite{ref:PDG} in order to improve the momentum resolution of the $\pi^0$.
To suppress backgrounds due to misreconstructed $\pi^0$ candidates, we require $|{\rm cos}\theta_{h}|<0.7$, where 
the helicity angle $\theta_h$ is defined as the angle between the direction of a $\gamma$ from a $\pi^{0}$ candidate in the $\pi^{0}$ rest-frame, 
and the $\pi^{0}$ direction in the laboratory.  

Photons from $\pi^0$ decays are a primary source of background in the region of the signal photon line from $h_b\rightarrow \gamma \eta_b$ 
transitions.
A signal photon candidate is rejected if, when combined with another photon
in the event ($\gamma_2$), the resulting $\gamma \gamma_2$ invariant mass is within 15 MeV/$c^2$ of the nominal $\pi^0$ mass; 
this is called a $\pi^0$ veto.  
Similarly, many misreconstructed $\pi^0$ candidates result from the pairing of photons from different 
$\pi^0$'s.  
A $\pi^0$ candidate is rejected  
if either of its daughter photons satisfies the $\pi^0$ veto condition, with $\gamma_2$ not the other daughter photon.
To maintain high signal efficiency, the $\pi^0$ veto condition is imposed only if the energy of $\gamma_2$ in the laboratory frame
is greater than $200$ MeV ($150$ MeV) for the signal photon (for the $\pi^0$ daughters).  
With the application of these vetoes, and after all selection criteria have been imposed, 
the average $\pi^0$ candidate multiplicity per event is 2.17 
for the full range of $m(\gamma \gamma)$, and 1.34 for the $\pi^0$ signal region ($110<m(\gamma \gamma)<150$ MeV/$c^2$).   
The average multiplicity for the signal photon is 1.02. For 
98.4\% of $\pi^0$ candidates there is only one associated photon candidate.  

We obtain the $m_\mathrm{recoil}(\pi^0)$ distribution in 90 intervals of 3 MeV/$c^2$ from 9.73 to 10 GeV/$c^2$. 
For each $m_\mathrm{recoil}(\pi^0)$ interval, the $m(\gamma \gamma)$ spectrum consists of a $\pi^0$ signal above combinatorial 
background (see Fig.~2). 
We construct the $m_\mathrm{recoil}(\pi^0)$ spectrum by extracting the $\pi^0$ signal yield in each interval of $m_\mathrm{recoil}(\pi^0)$ 
from a fit to the $m(\gamma\gamma)$ distribution in that interval.     
The $m_\mathrm{recoil}(\pi^0)$ distribution is thus obtained as the fitted $\pi^0$ yield and its uncertainty for each interval of $m_\mathrm{recoil}(\pi^0)$.

\begin{figure}[ht]
\begin{center}
 \includegraphics[width=.45\textwidth]{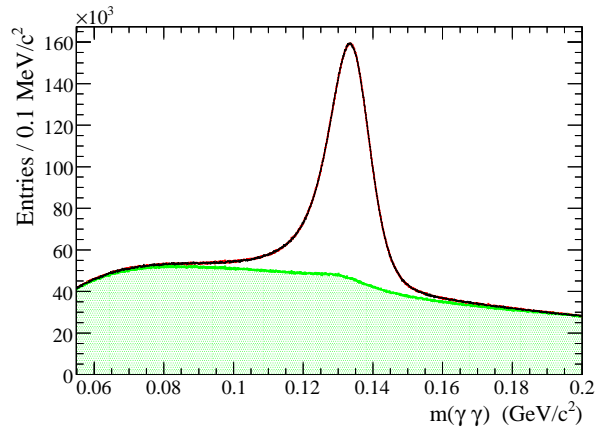}
 \begin{picture}(0.,0.)
    \end{picture}
  \caption{The result of the fit to the $m(\gamma \gamma)$ distribution in data (data points) for the full range of 
$m_\mathrm{recoil}(\pi^0$).  The solid 
histogram shows the fit result, and is essentially indistinguishable from the data; the shaded histogram
corresponds to the background distribution.  }
\end{center}
\end{figure}

We use the MC background and MC $\pi^0$-signal distributions 
directly in fitting the $m(\gamma\gamma)$ distributions in data~\cite{ref:ftn2}. 
For each $m_\mathrm{recoil}(\pi^0)$ interval in MC, we obtain histograms in 0.1 MeV/$c^2$ intervals of $m(\gamma \gamma)$ corresponding to 
the $\pi^0$-signal and background distributions.  The $\pi^0$-signal distribution is obtained by 
requiring matching of the reconstructed to the generated $\pi^0$'s on a candidate-by-candidate basis 
(termed ``truth-matching'' in the following discussion).  
The histogram representing background is obtained by subtraction of the $\pi^0$ signal from the total distribution. 

For both signal and 
background the qualitative changes in shape over the full range of $m_\mathrm{recoil}(\pi^0)$ are quite well reproduced 
by the MC.  
However, the $\pi^0$ signal distribution in data is slightly broader than in MC, and is peaked at a slightly higher mass value.    
The $m(\gamma \gamma)$  background shape also differs between data and MC.  
To address these differences, the MC $\pi^0$ signal is displaced in mass and smeared by a double Gaussian 
function with different mean and width values;  
the MC background distribution is weighted according to a polynomial in
$m(\gamma \gamma)$.  
The signal-shape and background-weighting parameter values are obtained from a fit to the $m(\gamma \gamma)$ distribution in data
for the full range of $m_\mathrm{recoil}(\pi^0)$.  
At each step
in the fitting procedure, the $\pi^0$ signal and background distributions
are normalized to unit area, and a $\chi^2$ between a linear combination of these MC histograms and the $m(\gamma \gamma)$ distribution in data is computed. 
The fit function provides an excellent description of the data ($\chi^{2}/NDF$=1446/1433; $NDF$=Number of Degrees of Freedom) 
and the fit result is essentially indistinguishable from the data histogram.  
The background distribution exhibits a small peak at the
$\pi^0$ mass, due to interactions in the detector material of
the type $n \pi^+ \rightarrow p \pi^0$ or $p \pi^- \rightarrow n \pi^0$ that cannot be truth-matched. The normalization of this
background to the non-peaking background is obtained from the MC simulation, which incorporates the
results of detailed studies of interactions in the detector material
performed using data~\cite{ref:geant}. This peak is displaced and smeared as for the
primary $\pi^0$ signal.

 The fits to the individual
$m(\gamma \gamma)$ distributions are performed with the smearing and weighting parameters fixed to the values
obtained from the fit shown in Fig.~2.  In this process, the MC signal and
background distributions for each $m_\mathrm{recoil}(\pi^0)$ interval are shifted,
smeared, and weighted using the fixed parameter values, and then
normalized to unit area.  Thus, only the signal and background yields are  free parameters
in each fit. 
The $\chi^2$ fit to the data then gives the value and
 the uncertainty of the number of $\pi^0$ events in each $m_\mathrm{recoil}$
 interval.  
The fits to the 90 $m(\gamma \gamma)$
distributions provide good descriptions of the data, with an average value of $\langle \chi^2/NDF\rangle = 0.98$ ($NDF$=1448), 
and r.m.s. deviation of 0.03 for the distribution of values.
We verify that the fitted $\pi^0$ yield is consistent with the number of truth-matched $\pi^0$'s in MC to ensure that the $\pi^0$ selection 
efficiency is well-determined, and to check the validity of the $\pi^0$ signal-extraction procedure.

To search for an $h_b$ signal, we perform a binned $\chi^{2}$ fit to the $m_\mathrm{recoil}(\pi^0$) distribution obtained in data.   
The $h_b$ signal function is represented by the sum of two 
Crystal Ball functions~\cite{ref:CB} with parameter values, other than the $h_b$ mass, $m$, and the normalization, 
 determined from simulated signal $\Upsilon(3S)\rightarrow \pi^0 h_b$ events.  
The background is well represented with a fifth order polynomial function.  

Direct MC simulation fails to yield an adequate description of the observed background distribution, although the overall shape is 
similar in data and MC.   
This is due primarily to the complete absence 
of experimental information on the decay modes of the $h_b$ and $\eta_b$ mesons.  
Simulation studies with a background component that is weighted
to accurately model the distribution in data show a negative bias of $\sim 35$\% in 
the signal yield from a procedure in which the background shape and signal mass 
and yield are determined simultaneously in the fit.  
Consequently, we define a region of $m_\mathrm{recoil}(\pi^0)$ 
chosen as the signal interval based on the expected mass value and signal resolution.  
The signal region includes any reasonable theoretical expectation for
 the $h_b$ mass.  
We fit the $m_\mathrm{recoil}(\pi^0)$ background distribution outside
the signal interval and interpolate the background to the signal region to obtain an estimate of
its uncertainty therein.  
Figure~3(a) shows the result of the fit to the distribution of 
$m_\mathrm{recoil}(\pi^0)$ in data excluding the signal region, $9.87\leq m_\mathrm{recoil}(\pi^0)\leq 9.93$ GeV/$c^2$. 
The fit yields  $\chi^2/NDF = 50.8/64$, and
the result is represented by the histogram in Fig.~3(a), including the interpolation to the $h_b$ signal region.  

We then perform a fit over the twenty intervals of the signal region to search for an $h_b$ signal of the expected
shape. 
We take account of the correlated uncertainties related to the polynomial interpolation procedure
 by creating a 20$\times$20 covariance matrix using the
6$\times$6 covariance matrix which results from the polynomial fit.  The error matrix for the signal region, $E$, is
obtained by adding the diagonal 20$\times$20 matrix of squared error values from the $m_\mathrm{recoil}(\pi^0)$ distribution,
and a $\chi^2$ value is defined by
\begin{eqnarray}
\chi^2=\tilde{V} E^{-1} V.
 \end{eqnarray}
Here $V$ is the column vector consisting of the difference between the measured value of the $m_\mathrm{recoil}(\pi^0)$
distribution and the corresponding sum of the value of the background polynomial and that of the $h_b$ signal 
function for each of the twenty 3 MeV/$c^2$ intervals
in the signal region. 
In Fig.~3(b) we plot the difference between the distribution of $m_\mathrm{recoil}(\pi^0)$ and the fitted histogram of 
Fig.~3(a) over the entire region from 9.73 GeV/$c^2$ to 10.00 GeV/$c^2$; we have combined pairs of 3 MeV/$c^2$ intervals from  
Fig.~3(a) for clarity.  
The yield obtained from the fit to the signal region is 10814$\pm$2813
events and the $h_b$ mass value obtained is $m = 9902 \pm 4$ MeV/$c^2$ with a $\chi^2$ value
of 14.7 for 18 degrees of freedom.

In order to determine the statistical significance of the signal we repeat
the fit with the $h_b$ mass fixed to the spin-weighted center of gravity of the $\chi_{bJ}(1P)$ states, $m=9900$ MeV/c$^2$.
The signal yield obtained from the fit is $10721\pm 2806$.  
The statistical significance 
of the signal, calculated from the square-root of the difference in $\chi^2$ for this fit
 with and without a signal component is 3.8 standard deviations, in good agreement with the signal size obtained.

Fit validation studies were performed.  No evidence of bias is observed in 
 large MC samples with simulated $h_b$ mass at 9880, 9900, and 9920 MeV/$c^2$.                                       
In addition, the result of a scan performed in data as a function of the assumed $h_b$ mass indicates that the preferred peak position
for the signal is at 9900 MeV/$c^2$, in excellent agreement with the result of Fig.~3(b).

\begin{figure}[!ht]
\begin{center}
\includegraphics[width=.45\textwidth]{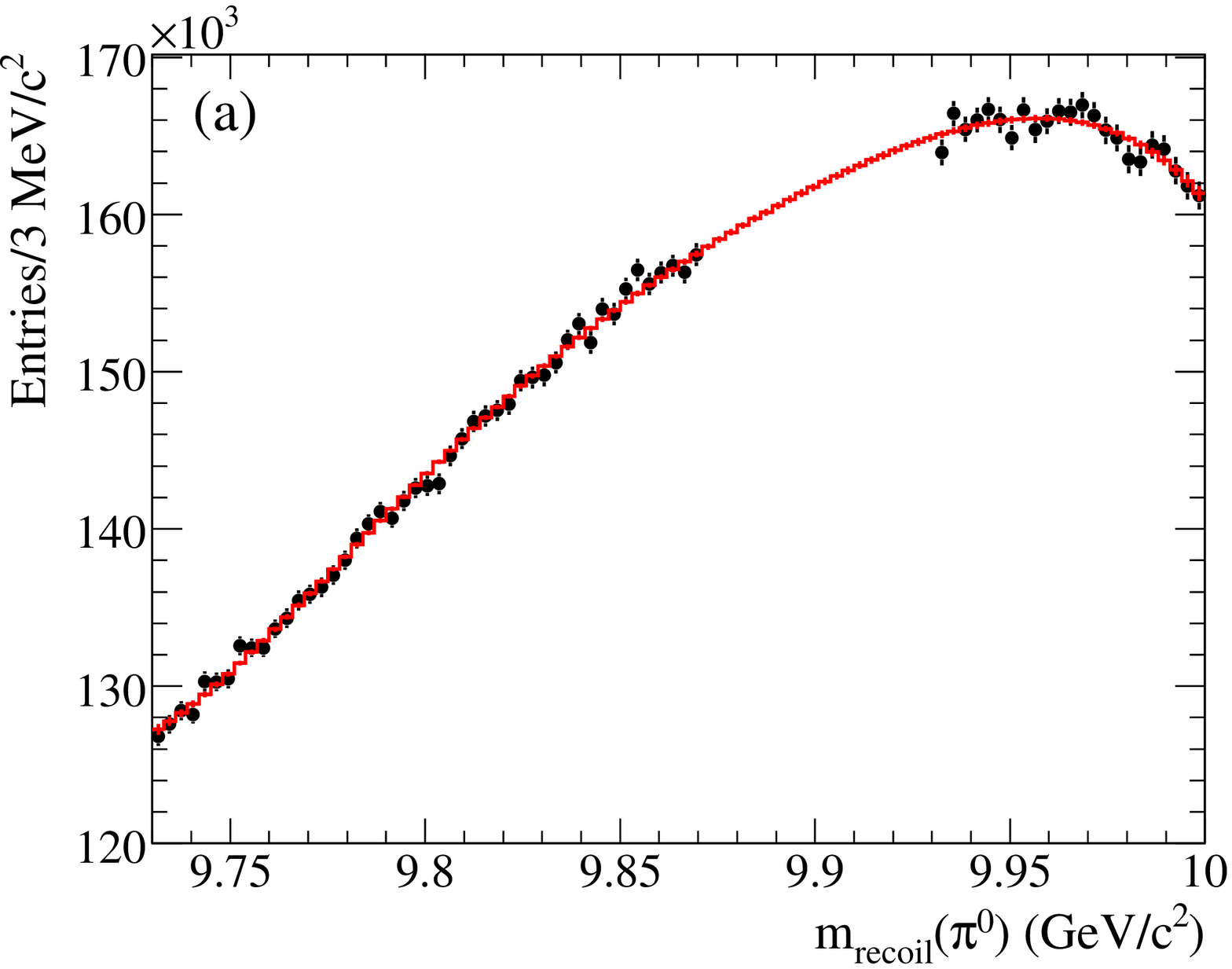}
\includegraphics[width=.45\textwidth]{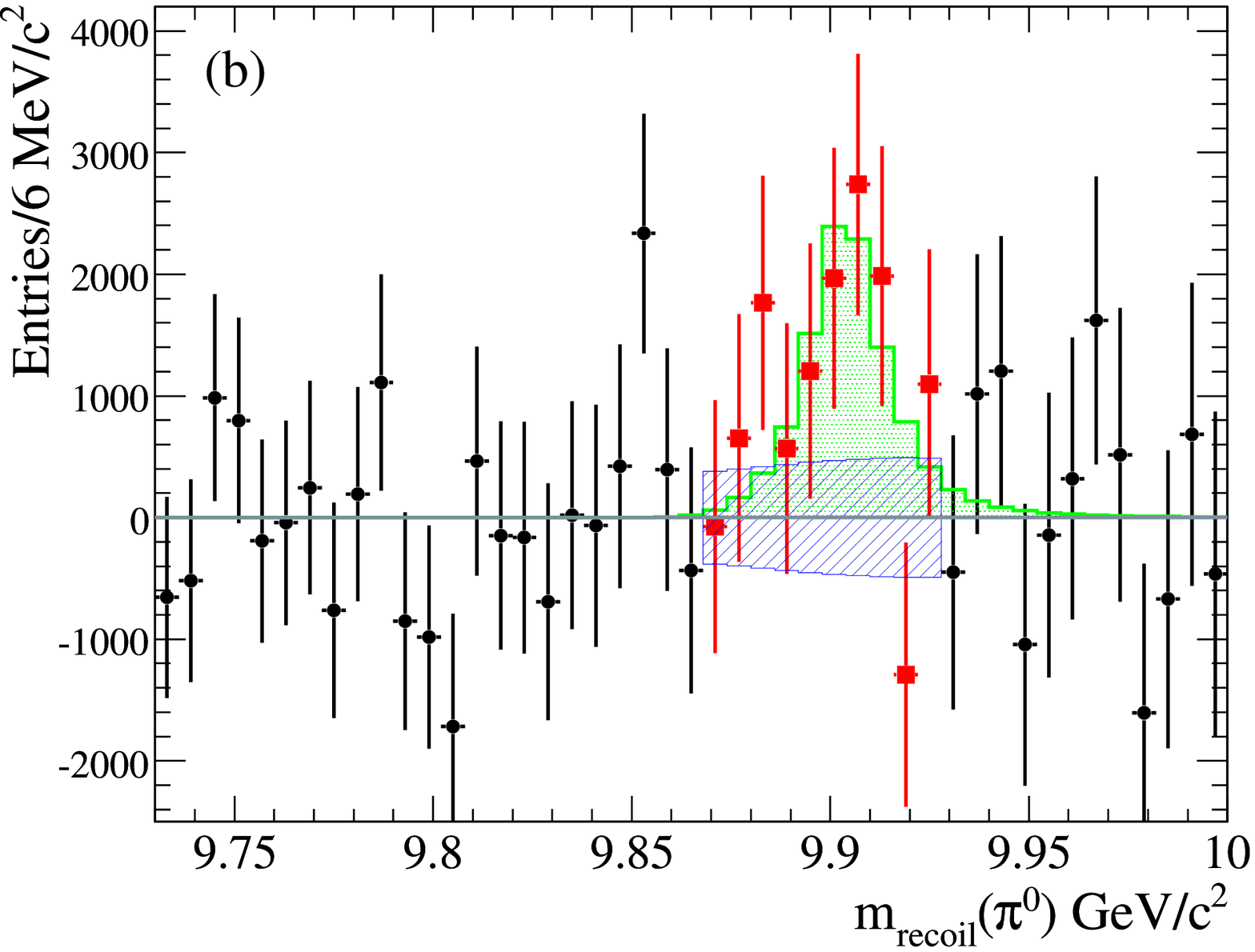}
  \caption{ (a) The $m_\mathrm{recoil}(\pi^0$) distribution in the region $9.73<m_\mathrm{recoil}(\pi^0)<10$ GeV/$c^2$ for data (points); 
the solid histogram represents the fit function described in the text.  
The data in the $h_b$ signal region have been excluded from the fit and the plot.  
(b) The $m_\mathrm{recoil}(\pi^0$) spectrum after subtracting background; in the $h_b$ signal region the data points are shown as squares, 
and the area with diagonal shading represents the uncertainties from the background fit;  
the shaded histogram represents the signal function resulting from the fit to the 
data.}
\end{center}
\end{figure}

We obtain an estimate of systematic uncertainty on the number of $\pi^0$'s in each $m_\mathrm{recoil}(\pi^0)$ interval 
by repeating the fits to the individual $m(\gamma\gamma)$ 
spectra with the lineshape parameters corresponding to Fig.~2 varied within their uncertainties. 
The distribution of the net uncertainty varies as a third order polynomial in $m_\mathrm{recoil}(\pi^0$).  
We estimate a systematic uncertainty of $\pm$210 events on the $h_b$ signal yield due to the $\pi^0$-yield extraction procedure 
by evaluating this function at the fitted $h_b$ mass value.

The dominant sources of systematic uncertainty on the measured $h_b$ yield 
are the order of the polynomial describing the $m_\mathrm{recoil}(\pi^0$) background
distribution, and the width of the $h_b$ signal region.  
By varying the polynomial from fifth- to seventh-order, and by expanding the region excluded 
from the fit in Fig.~3(a) from (9.87--9.93) GeV/$c^{2}$ to (9.85--9.95) GeV/$c^{2}$,   
we obtain systematic uncertainties of $\pm 1065$ events and $\pm 1263$ events, respectively,  
taken from the full excursions of the $h_b$ yield under these changes.  
Similarly, we obtain a total systematic uncertainty of $\pm 1.5$ MeV/$c^2$ on the $h_b$ mass due to the choice of background shape.

The systematic uncertainty associated with the choice of signal lineshape is estimated by varying 
the signal function parameters, which were fixed in the fit, by $\pm 1\sigma$.  We assign the largest deviation from the 
nominal fit result as a systematic error. 
Systematic uncertainties of $\pm 154$ events and $\pm 0.3$ MeV/$c^2$ are obtained for the $h_b$ yield and mass, respectively.

After combining these systematic uncertainty estimates in quadrature, we obtain an effective signal significance 
of 3.3 standard deviations.  The smallest value of
the significance among those calculated for the varied fits in the
systematics study is 3.1 standard deviations.  The 
$h_b$ yield is $10814\pm 2813 \pm 1652$ events and the $h_b$ 
mass value $m=9902\pm 4 \pm 2$ MeV/$c^2$, where the first uncertainty is statistical and the second systematic.  
The resulting hyperfine splitting with respect to the center of gravity of the $\chi_{bJ}(1P)$ states is thus $\Delta M_{\rm HF} =+2\pm 4 \pm 2$ MeV/$c^2$, 
which agrees within error with model predictions~\cite{ref:BFpredict,ref:Godfrey}. 

To convert the $h_b$ signal yield into a measurement of the product BF for the sequential decay $\Upsilon(3S)\rightarrow \pi^0 h_b$, $h_b\rightarrow \gamma \eta_b$, 
we determine the efficiency $\epsilon_S$ from MC by requiring that the signal $\pi^0$ and the $\gamma$ 
be truth-matched.  The resulting efficiency is $\epsilon_S$ = $15.8\pm 0.2$\%.
Monte Carlo studies indicate that photons that are not from an $h_b\rightarrow \gamma \eta_b$ transition can satisfy the selection criteria  
when only the $\Upsilon(3S)\rightarrow \pi^0 h_b$ transition is truth-matched.  This causes a fictitious 
increase in the $h_b$ signal efficiency to $\epsilon$ = 17.9$\pm$ 0.2\%.  
Therefore, the efficiency for observed $h_b$ signal events that do not correspond to $h_b\rightarrow \gamma \eta_b$ decay 
is $\Delta\epsilon$ = 2.1\%.  
However, there is no current experimental information on the production of such non-signal photons in $h_b$
and $\eta_b$ decays.  Furthermore, the above estimate of efficiencies in MC does not account for photons from 
hadronic $h_b$ decays, since the signal MC requires $h_b\rightarrow \gamma \eta_b$.  We thus assume that random photons 
from hadronic $h_b$ decays have the same probability $\Delta\epsilon$ to satisfy the signal photon selection criteria as those from $\eta_b$ decays. 
We assume a 100\% uncertainty on the value of $\Delta\epsilon$ when estimating the systematic error on the product BF.  

We estimate the product BF for 
$\Upsilon(3S)\rightarrow \pi^0 h_b$, $h_b\rightarrow \gamma \eta_b$ by dividing the fitted signal yield, $N$, corrected for the estimated total 
reconstruction efficiency, by the number of
$\Upsilon(3S)$ events, $N_{\Upsilon(3S)}$, in the data sample. 
We obtain the following expression for the product BF:
\begin{eqnarray}
  {\mathcal B}(\Upsilon(3S)\rightarrow \pi^0 h_b)\times {\mathcal B}(h_b\rightarrow \gamma \eta_b) 
 = \frac{N}{N_{\Upsilon(3S)}\, \epsilon_S}\cdot\frac{1}{C}, 
 \end{eqnarray}
where 
\begin{eqnarray}
C=1+\frac{\Delta\epsilon}{\epsilon_S}\cdot\frac{1}{{\mathcal B}(h_b\rightarrow \gamma \eta_b)}
\end{eqnarray}
is the factor that corrects the efficiency $\epsilon_S$ for the non-signal hadronic $h_b$ and $\eta_b$ contributions.  
In this equation, we assume a BF value ${\mathcal B}(h_b\rightarrow \gamma \eta_b)=45\pm 5$\% according to the current range of theoretical predictions.   
The corresponding correction factor is $1-C\sim 30$\%, with a systematic uncertainty dominated by the uncertainty on $\Delta\epsilon$.  

We obtain ${\mathcal B}(\Upsilon(3S)\rightarrow \pi^0 h_b)\times {\mathcal B}(h_b\rightarrow \gamma \eta_b)=(4.3 \pm 1.1 \pm 0.9) \times 10^{-4}$, where
the first uncertainty is statistical
and the second systematic.
The result is consistent with the prediction of Ref.~\cite{ref:Godfrey}, which estimates $4\times 10^{-4}$ for the product BF.
Since the $h_b$-decay uncertainty reduces the significance of the product BF relative to that of the $h_b$ production, we may 
also quote an upper limit on the product BF.  
From an ensemble of simulated events using the measured product BF value, and the statistical and associated systematic uncertainties
(assumed to be Gaussian) as input, we obtain 
${\mathcal B}(\Upsilon(3S)\rightarrow \pi^0 h_b)\times {\mathcal B}(h_b\rightarrow \gamma \eta_b)<6.1\times 10^{-4}$ at 90\% C.L.

In summary, we have found evidence for the decay $\Upsilon(3S)\rightarrow \pi^0 h_b$, 
with a significance of at least 3.1 standard deviations, including systematic uncertainties. 
The measured mass value, $m=9902\pm 4$(stat.)$\pm 2$(syst.) MeV/$c^2$,  
 is consistent with the expectation for the $h_b(1P)$ bottomonium state~\cite{ref:Rosner2002, ref:Meinel}, 
the axial vector partner of the $\chi_{bJ}(1P)$ triplet of states.
We obtain 
${\mathcal B}(\Upsilon(3S)\rightarrow \pi^0 h_b)\times {\mathcal B}(h_b\rightarrow \gamma \eta_b)=(4.3 \pm 1.1$ (stat.)$\pm 0.9$ (syst.))$\times 10^{-4}$ 
($<6.1\times 10^{-4}$ at 90\% C.L.).

\vspace{.1 in}

\begin{acknowledgments}
We are grateful for the excellent luminosity and machine conditions
provided by our \pep2\ colleagues, 
and for the substantial dedicated effort from
the computing organizations that support \babar.
The collaborating institutions wish to thank 
SLAC for its support and kind hospitality. 
This work is supported by
DOE
and NSF (USA),
NSERC (Canada),
CEA and
CNRS-IN2P3
(France),
BMBF and DFG
(Germany),
INFN (Italy),
FOM (The Netherlands),
NFR (Norway),
MES (Russia),
MICIIN (Spain),
STFC (United Kingdom). 
Individuals have received support from the
Marie Curie EIF (European Union),
the A.~P.~Sloan Foundation (USA)
and the Binational Science Foundation (USA-Israel).
\end{acknowledgments}

\vspace{.1 in} 

{\bf Note added in proof:} After this paper was submitted, preliminary results of a search for the $h_b$ in the reaction 
$e^+e^-\rightarrow h_b(nP) \pi^+\pi^-$ in data collected near the $\Upsilon(5S)$ resonance 
have been announced by the Belle Collaboration~\cite{ref:Belle}. The $h_b(1P)$ mass measured therein agrees very well with 
the value reported in this paper.   

\vspace{2.5 in}
 

\begin{thebibliography} {19}
\bibitem{ref:ftn1}
{The MC events are generated using the \jetset74 and 
PYTHIA programs to describe the hadronization process from the Lund
string fragmentation model with final-state radiation included.}
\bibitem{ref:r2} This quantity is indicative of the collimation of an event topology, with values close to one for jetlike events;
the kinematics of a heavy object such as the $\eta_b$ decaying hadronically result in a more spherical event.
\bibitem{ref:ftn2}
{In MC simulations, fits to the individual $m(\gamma \gamma)$ spectra that make use of a polynomial background 
function and various combinations of Crystal Ball~\protect\cite{ref:CB} and/or Gaussian signal functions 
proved unsatisfactory at the high statistical precision necessary.}
\end{thebibliography}
\end{document}